\documentclass[a4paper,fleqn,usenatbib]{mnras}
\usepackage[T1]{fontenc}
\usepackage{ae,aecompl}

\usepackage{graphicx}   
\usepackage{amsmath}    
\usepackage{amssymb}    

\newcommand{\figureref}[1]{Fig.~\ref{#1}}

\newcommand{\multifigureref}[2]{Fig.~\ref{#1} and Fig.~\ref{#2}}

\newcommand{\tableref}[1]{Table~\ref{#1}}

\newcommand{\sectionref}[1]{Section \ref{#1}}

\DeclareMathOperator{\sech}{sech}

\title[RHD simulations of the DIG]{Radiation hydrodynamics simulations 
of the evolution of the diffuse ionized gas in disc galaxies}

\author[B. Vandenbroucke \& K. Wood]{
Bert Vandenbroucke,$^{1}$\thanks{E-mail~: bv7@st-andrews.ac.uk}
Kenneth Wood,$^{1}$
\\
$^1$SUPA, School of Physics \& Astronomy, University of St Andrews, North Haugh,
St Andrews, KY16 9SS, United Kingdom
}

\date{Accepted XXX. Received YYY; in original form ZZZ}

\pubyear{2019}

\begin{document}

\label{firstpage}
\pagerange{\pageref{firstpage}--\pageref{lastpage}}
\maketitle

\begin{abstract}

There is strong evidence that the diffuse ionized gas (DIG) in disc 
galaxies is photoionized by radiation from UV luminous O and B stars in 
the galactic disc, both from observations and detailed numerical models. 
However, it is still not clear what mechanism is responsible for 
providing the necessary pressure support for a diffuse gas layer at 
kpc-scale above the disc. In this work we investigate if the pressure 
increase caused by photoionization can provide this support. We run 
self-consistent radiation hydrodynamics models of a gaseous disc in an 
external potential. We find that photoionization feedback can drive low 
levels of turbulence in the dense galactic disc, and that it provides 
pressure support for an extended diffuse gas layer. Our results show 
that there is a natural fine-tuning between the total ionizing radiation 
budget of the sources in the galaxy and the amount of gas in the 
different ionization phases of the ISM, and provide the first fully 
consistent radiation hydrodynamics model of the DIG.

\end{abstract}


\section{Introduction}

The existence of a diffuse layer of ionized material in our Galaxy was 
first proposed by \citet{1963Hoyle} to explain an observed absorption 
feature in the galactic synchrotron background. The existence of this 
layer was subsequently confirmed through measurements of its faint 
H$\alpha{}$ emission \citep{1973Reynolds}; currently large galaxy-wide 
surveys are available that provide a full picture of what is now 
reffered to as the Warm Ionized Medium (WIM) \citep{2003Haffner}. 
\citet{1990Dettmar} and \citet{1990Rand} found similar diffuse ionized 
layers in other galaxies, and introduced the general term \emph{diffuse 
ionized gas} (DIG). The spatial distribution and properties of the DIG 
have been extensively studied; see the review by \citet{2009Haffner}. 
The DIG corresponds to the so called \emph{warm} phase of the ISM; the 
dense disc constitutes a \emph{cold} phase, while a fraction of the gas 
is expected to be in a \emph{hot} phase at lower densities and higher 
temperatures than the DIG \citep{1974Cox, 1977McKee}. For this work, we 
will only consider the cold and warm phases of the ISM.

To explain the existence of the DIG, we need to address two important questions:
\begin{enumerate}
  \item{} What mechanism is responsible for ionizing the diffuse gas?
  \item{} What mechanism is responsible for supporting a diffuse gas layer at
  the observed heights above the galactic disc?
\end{enumerate}

An extensive body of literature exists that addresses the first 
question. The observed ionization fraction of the DIG can only be 
explained if the DIG is photoionized by local sources, i.e. UV luminous 
O and B stars in the galactic disc \citep{1995Reynolds}. Furthermore, 
there are strong correlations between the presence of a high star 
formation rate and the signal from DIG emission \citep{2003Rossa}, and 
between line emission from {\sc{}Hii} regions surrounding young O stars 
and DIG emission \citep{1996Ferguson_HIIregions}. To explain how the UV 
radiation from these sources makes it out of the dense galactic disc to 
the large altitudes where the DIG is observed, we have to assume a 
turbulent disc structure that contains low density channels 
(\emph{chimneys}) through which ionizing radiation can escape 
\citep{2009Haffner}. Post-processing of realistic disc models 
\citep{2006Joung, 2012Hill, 2016Girichidis_CR} has shown that these 
channels do indeed exist \citep[from here on V18]{2010Wood, 2014Barnes, 
2018Vandenbroucke_SILCC}.

The temperature of the DIG is typically higher than that in {\sc{Hii}} 
regions in the disc, and tends to increase with altitude above the disc, 
both in our Galaxy \citep{1999Haffner, 2006Madsen} and other galaxies 
\citep{1996Ferguson_NGC55}. \citet{2004Wood_spectralhardening} showed 
that such a temperature structure can be explained by spectral hardening 
of the radiation field, which is a possible scenario if the sources of 
ionizing radiation are embedded within a dense molecular cloud and the 
DIG is ionized by so called \emph{leaky} \textsc{Hii} regions 
\citep{2003Hoopes}. V18 showed that spectral hardening does indeed 
produce the expected line emission signature in post-processed disc 
galaxy models, provided that the UV luminosity of the ionizing sources 
is fine tuned to reproduce both the scale height of the DIG (as measured 
from the scale height of the ${\rm{}H}\alpha{}$ profile) and the 
temperature structure. In these models, the spectral hardening occurs 
inside the DIG and it is the spectral change throughout the DIG that 
reproduces the observed increase in line emission. Note that the 
observed temperature structure of the disc can also be explained by 
considering the impact of hot evolved stars \citep{2011Rand, 
2011FloresFajardo}, and that a full model for the DIG should also take 
into account other sources of UV radiation.

Models of the DIG until now have failed to address the second question. 
Those attempts that post-processed realistic disc galaxy models were 
dependent on the specific model feedback implementations to explain the 
presence of diffuse gas in these models; when a diffuse component is 
present, it can be ionized and produces a realistic DIG. However, models 
without sufficient diffuse gas at high altitudes are less satisfactory 
\citep[V18]{2014Barnes}. Clearly, some mechanism is 
required to provide pressure support for diffuse gas at high altitudes.

In this paper, we build on the correlation found between the ionizing 
source luminosity and the scale height of the DIG in V18, and propose 
that photoionization itself is responsible for producing the necessary 
pressure support to maintain the DIG. Neutral atomic gas has a typical 
temperature of $100-500~{\rm{}K}$, while gas in the DIG has an 
equilibrium temperature of $\sim{}8,000~{\rm{}K}$ and more. This 
increase in temperature gives the DIG a significantly higher 
hydrodynamic pressure, which provides extra support against the 
gravitational potential of the disc. We know that the ionizing radiation 
has to make it out to the high altitudes where the DIG is supported, so 
this mechanism will be effective at these altitudes.

To quantify this effect, we will run idealized radiation hydrodynamics 
(RHD) simulations of a patch of a disc galaxy that is initially in 
ionized hydrostatic equilibrium in an external disc potential. We will 
then add a realistic distribution of luminous O stars to this setup and 
study how the UV radiation of these sources affects the dynamics of the 
gas in the disc. We post-process these simulations using the same model 
as employed in V18 (and using the same number of sources and source 
positions as in the corresponding RHD model), and verify that this still 
produces a DIG that is in line with the observed DIG. To our knowledge, 
this is the first self-consistent RHD study of the DIG; previous models 
of ionized gas in disc galaxies focussed on the central galactic disc 
\citep{2012deAvillez, 2017Seifried} and did not study the impact of 
radiation on the extended diffuse disc.

The paper is structured as follows: first, we will introduce our method. 
We will then show that our models give converged results, and discuss 
the effect of our model parameters on the results. We will end with our 
conclusions.

\section{Method}

Full radiation transfer modelling of the DIG and all relevant cooling 
and heating mechanisms is computationally expensive. We will therefore 
employ a two-step technique to model the dynamics of the DIG and the 
resulting DIG structure:
\begin{itemize}
  \item{} first, we will run RHD models in which the radiation field is
  treated using a simplified two-temperature model with hydrogen as the
  only absorber of ionizing radiation \citep[Falceta-Gon\c{c}alves \emph{et al.},
  \emph{in prep.}]{2019Lund},
  \item{} then, we will post-process these simulations to produce a more
  complete model of the temperature structure and ionization state, using the
  full V18 model. Note that these models are only intended to
  check that the spectral hardening found in V18 is still effective, and not
  to produce a full spectral model for the DIG.
\end{itemize}

For both steps, we will use the Monte Carlo RHD code 
\textsc{CMacIonize}\footnote{\url{https://github.com/bwvdnbro/CMacIonize}} 
\citep{2018Vandenbroucke_CMacCode, 2018Vandenbroucke_CMacIonize}, both 
in its full RHD mode and its post-processing mode. For consistency, we 
will use the same grid and source model for the RHD step and the 
post-processing step.

Below we discuss both models and our initial conditions in more detail.

\subsection{RHD model}

\subsubsection{External potential}

We will assume that the total mass contained in the gas in our 
simulations is only a small fraction (10\%) of the total mass budget, 
so that we will approximate the gravitational potential with an 
external, analytic potential. To this end, we use the same isothermal 
sheet potential used by the SILCC project \citep{2015Walch}, with the 
gravitational acceleration $\boldsymbol{a}_{\rm{}grav}$ given by
\begin{equation}
\boldsymbol{a}_{\rm{}grav} = - 2 \pi{} G \Sigma{}_M \tanh\left(
\frac{z}{b_M}\right),
\end{equation}
with $z$ the altitude above the plane of the disc and $G = 
6.674\times{}10^{-11}~{\rm{}m}^3~{\rm{}kg}^{-1}~{\rm{}s}^{-2}$ Newton's 
constant. The two parameters for the potential are the surface density 
$\Sigma{}_M$ and the scale height $b_M$, for which we will use the same 
values as used for SILCC: $\Sigma{}_M = 
30~{\rm{}M}_\odot{}~{\rm{}pc}^{-2}$ and $b_M = 200~{\rm{}pc}$.

\citet{2013Creasey} parametrise the isothermal sheet potential in terms 
of the gas surface density $\Sigma{}_g = f_g \Sigma{}_M$, with $f_g$ the 
fraction of the total mass that is in gas (we assume $f_g = 0.1$). 
Assuming that the gas has the same density distribution as the other 
mass components, they find an expression for the gas scale height $b_g 
(=b_M)$ for gas in hydrostatic equilibrium with the potential at 
temperature $T$:
\begin{equation}
b_g = \frac{f_g k_B T}{\mu{} m_p \pi{} G \Sigma{}_g}
\approx{} \frac{611}{\mu{}} \left(\frac{T}{1000~{\rm{}K}}\right)
\left(\frac{\Sigma{}_M}{1~{\rm{}M}_\odot{}~{\rm{}pc}^{-2}}
\right)^{-1}~{\rm{}pc},
\end{equation}
with $k_B = 1.38064852\times{}10^{-23}~{\rm{}J~K}^{-1}$ Boltzmann's 
constant, $m_p = 1.6726219\times{}10^{-27}~{\rm{}kg}$ the mass of a 
proton and $\mu{}$ the mean molecular weight of the gas. The SILCC 
choice of parameters hence corresponds to a neutral hydrogen only gas at 
$T \approx{} 10^4~{\rm{}K}$.

In \ref{subsection:initial_conditions} we will discuss the density 
distribution that corresponds to this potential.

\subsubsection{Equation of state}

We will assume a two-temperature isothermal equation of state for the 
gas, with the pressure $P$ given by
\begin{equation}
P = \frac{\rho{} k_B T_{\rm{}eq}}{\mu{}_{\rm{eq}} m_p},
\end{equation}
with $\rho{}$ the gas density and $T_{\rm{}eq}$ and $\mu{}_{\rm{}eq}$ 
the equilibrium temperature and mean molecular mass, given by
\begin{equation}
T_{\rm{}eq} = x_{\rm{H}} T_n + (1 - x_{\rm{}H}) T_i,
\end{equation}
and
\begin{equation}
\mu{}_{\rm{}eq} = x_{\rm{}H} + \frac{1}{2} (1 - x_{\rm{}H}) =
\frac{1}{2} (x_{\rm{}H} + 1),
\end{equation}
with $x_{\rm{}H}$ the neutral fraction of hydrogen for a cell. The two 
temperatures $T_n$ and $T_i$ correspond to the assumed equilibrium 
temperatures for neutral and ionized gas respectively. We will assume 
$T_n=500~{\rm{}K}$ and $T_i=8,000~{\rm{}K}$. Note that we choose a 
relatively high neutral gas temperature to account for the fact that we 
do not include any additional feedback processes: at lower temperatures 
our density profile becomes very steep and it is very hard for the 
radiation to create outflow cavities through which ionizing radiation 
escapes into the diffuse disc. In the presence of additional feedback 
mechanisms, we would expect these cavities to be present and facilitate 
the escape of ionizing radiation.

\subsubsection{Radiation}
\label{subsubsection:radiation}

\begin{figure}
\centering{}
\includegraphics[width=0.48\textwidth]{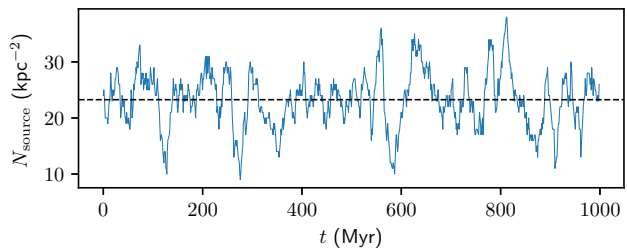}
\caption{Number of ionizing sources, $N_{\rm{}source}$, as a function of time for one of our
simulations. Note that our source model ensures that this curve is the same for
all simulations. The black dashed line indicates the integrated average.
\label{figure:number_of_sources}}
\end{figure}

To simplify the radiation transfer during the RHD simulations, we will 
assume a monochromatic ionizing spectrum with an energy of 
$13.6~{\rm{}eV}$, and a constant photoionization cross section of 
$6.3\times{}10^{-18}~{\rm{}cm}^2$ (since the photon energy only affects 
its photoionizing cross section in this problem, the use of a 
monochromatic spectrum is not strictly necessary). We furthermore assume 
a constant recombination rate of 
$2.7\times{}10^{-13}~{\rm{}cm}^3~{\rm{}s}^{-1}$. This implies a 
recombination time scale $\Delta{}t_R = 1.2\times{}10^5 \left(n_{\rm{}H} 
/ {\rm{}cm}^{-3} \right)^{-1}~{\rm{}yr}$. Depending on the resolution, 
we use $10^7$ or $10^8$ photon packets per iteration for 10 iterations. 
To limit the impact of the radiation step on the integration, we do not 
do the full radiative transfer calculation after every hydrodynamical 
integration step, but instead update the ionization state on a regular 
time interval basis. We run simulations with radiation intervals of 
$0.1~{\rm{}Myr}$, $0.5~{\rm{}Myr}$ and $1~{\rm{}Myr}$. We will show the 
effect this has on our final result below.

As an ionizing source model, we will use the same distribution as used 
in V18, with uniform random coordinates in the plane of the disc, and a 
Gaussian vertical distribution perpendicular to the disc, with a scale 
height of $63~{\rm{}pc}$. We assume an average source number density of 
$24~{\rm{}kpc}^{-2}$ and an average source life time of $20~{\rm{}Myr}$. 
The ionizing luminosity per source, $Q_{\rm{}H}$, is a parameter in our 
model; we will explore three different values: $10^{47}~{\rm{}s}^{-1}$, 
$10^{48}~{\rm{}s}^{-1}$ and $10^{49}~{\rm{}s}^{-1}$. These values 
correspond to respectively 1~\%, 10~\% and 100~\% of the ionizing 
luminosity of an average O/B star, and take into account the fact that 
we do not resolve the dense birth clump of these stars which can be 
expected to absorb a significant fraction of the ionizing radiation. To 
limit the number of parameters in our model, we assume the same ionizing 
luminosity per source and ignore variations because of stochastic 
sampling of the IMF.

We keep track of the life time of each individual source, and allow for 
the creation and destruction of sources over time. To this end, we start 
by sampling 24 source positions at the start of the simulation, and give 
each source a life time sampled from a uniform distribution in $[0, 
20]~{\rm{}Myr}$. At regular intervals (we use a value $1~{\rm{}Myr}$ 
corresponding to the longest time between radiation steps in all our 
simulations), we remove sources whose life time was exceeded (we 
currently ignore any impact of supernova feedback at the end of a source 
lifetime). We then check for the potential creation of new sources: we 
draw 24 random numbers and for each of them check if they are larger or 
smaller than the source creation probability $(1~{\rm{}Myr}) / 
(20~{\rm{}Myr}) = 5~\%$. For each random number that is smaller, we 
create an additional source with a life time of $20~{\rm{}Myr}$ that is 
uniformly offset within the $1~{\rm{}Myr}$ interval. 
\figureref{figure:number_of_sources} shows the resulting number of 
sources as a function of time. Note that we keep the source distribution 
fixed for all our models and only vary the ionizing luminosity of each 
individual source.

Note that we do only include the effect of sources in the disc, and do 
not consider the impact of hot evolved stars, as done in e.g. 
\citet{2011Rand, 2011FloresFajardo}. The reason for this is that the 
total ionizing luminosity of O/B stars is significantly higher than that 
of hot evolved stars: \citet{2011FloresFajardo} find an ionizing 
luminosity $Q_{\rm{}H}=7.8\times{}10^{53}~{\rm{}s}^{-1}$ for the former 
and $Q_{\rm{}H}=2.1\times{}10^{51}~{\rm{}s}^{-1}$ for the latter. The 
ionizing luminosity of hot evolved stars is hence only a few percent of 
that of the more luminous sources in the disc, even if only 
$\sim{}$10~\% of the O/B star luminosity makes it out of the unresolved 
birth cloud surrounding it. Additionally, the O/B star luminosity is 
significantly more concentrated in space. The appearance and 
disappareance of a small number of O/B stars will hence have a much more 
significant dynamical impact than the gradual changes in a diffuse and 
more extended distribution of hot evolved stars.

Despite their relatively small dynamic effect, hot evolved stars will 
contribute to the heating of the DIG, as the harder spectrum of these 
sources can penetrate the diffuse gas more easily. A fully 
self-consistent model for the spectral emission from the DIG should 
therefore include these sources; we choose to focus on the impact of O/B 
stars alone in this work.

\subsubsection{Simulations}

We sample a slice of the disc in a box of 
$1\times{}1\times{}8~{\rm{}kpc}$ with periodic boundaries in the $x$ and 
$y$ directions and semi-permeable boundaries in the $z$ direction (we 
allow material to leave the box, but we do not allow material to enter). 
We find that these semi-permeable boundaries are necessary to prevent 
unrealistically large inflows through the boundaries. We run simulations 
at two different resolutions: $64\times{}64\times{}512$ and 
$128\times{}128\times{}1024$ cells, corresponding to a spatial 
resolution of $15.6~{\rm{}pc}$ and $7.8~{\rm{}pc}$ respectively. Due to 
computational limitations, we will only run one high resolution 
simulation and use that to assess the impact of resolution; we will not 
attempt to obtain spatially converged results.

All simulations were run for a total time of at least $800~{\rm{}Myr}$, 
with a global time step set by the hydrodynamical stability condition 
(we use a CFL factor of 0.2). The high resolution simulation was only 
run for $200~{\rm{}Myr}$. The hydrodynamics scheme is the same as 
described in \citet{2018Vandenbroucke_CMacIonize}, with some minor 
modifications to deal with numerical issues resulting from the very low 
densities at high altitudes above the disc in the presence of an 
external acceleration due to the gravitational potential: a conservative 
flux limiter that suppresses unrealistically large gravitationally 
driven mass fluxes, and a velocity limiter that artificially limits the 
fluid velocity and sound speed to $1,000~{\rm{}km~s}^{-1}$. We find that 
these modifications have no noticeable impact on the result of our 
simulations, but are necessary to maintain stability. The hydrodynamical 
fluxes are computed using an approximate HLLC Riemann solver.

\subsection{Initial conditions}
\label{subsection:initial_conditions}

As initial conditions, we assume an ionized hydrogen gas 
($T=T_i=8,000~{\rm{}K}$) in hydrostatic equilibrium with the external 
potential. Since our initial gas temperature is not the same as the 
assumed equilibrium temperature used to determine the scale height for 
the potential, we cannot use the density profile from 
\citet{2013Creasey}, but instead we need to use the more general 
expression
\begin{equation}
\rho{} = \frac{\Sigma{}_g'}{2b_M} \left(\sech\left(\frac{z}{b_M}\right)
\right)^{\frac{2b_M}{b_g}}.
\end{equation}
In the specific case $b_g = b_M$ we recover the density profile used by
\citet{2013Creasey}.

The gas surface density $\Sigma{}_g'$ now has the more general form
\begin{equation}
\Sigma{}_g' = I(b_M / b_g) f_g \Sigma{}_M,
\end{equation}
with
\begin{equation}
I(x) = 2 \left(\int_{-\infty{}}^{+\infty{}} \sech^{2x} (y) ~{\rm{}d}y
\right)^{-1}.
\end{equation}
This expression can be derived by imposing $M_g = f_g M_M$, where $M_g$ and
$M_M$ are the total gas mass and total mass within an infinitely high box.

\begin{table}
\centering{}
\caption{Values for the fit parameters in \figureref{figure:mass_integral}.
\label{table:mass_integral}}
\begin{tabular}{c c}
\hline{}
$A$ & $0.015$ \\
$B$ & $-0.085$ \\
$C$ & $0.635$ \\
$D$ & $-0.010$ \\
\hline{}
\end{tabular}
\end{table}

\begin{figure}
\centering{}
\includegraphics[width=0.48\textwidth]{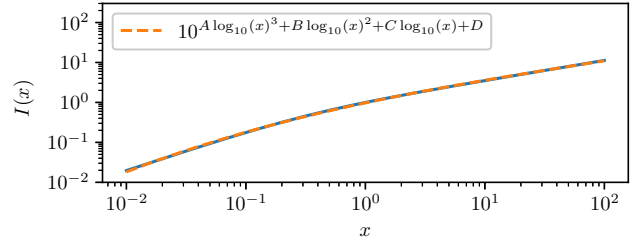}
\caption{Value of the function $I(x)$ for a realistic range of scale height
ratios $x$. The full line represents the numerically evaluated value, while
the dashed line is a 4 parameter fit, as indicated in the legend.
\label{figure:mass_integral}}
\end{figure}

The integral $I(x)$ has no general analytic solution. We compute it 
numerically for a large range of $x$ values and interpolate on these to 
get its value for arbitrary scale height ratios. The value of $I(x)$ and 
our interpolating curve are shown in \figureref{figure:mass_integral} 
and \tableref{table:mass_integral}. For small scale height ratios ($b_g$ 
larger than $b_M$) the gas mass is more spread out than the total mass 
and the gas surface density is lower than $f_g\Sigma{}_M$. For large 
ratios (small $b_g$), the surface density is higher, reflecting a more 
concentrated gas density profile.

\begin{figure}
\centering{}
\includegraphics[width=0.48\textwidth]{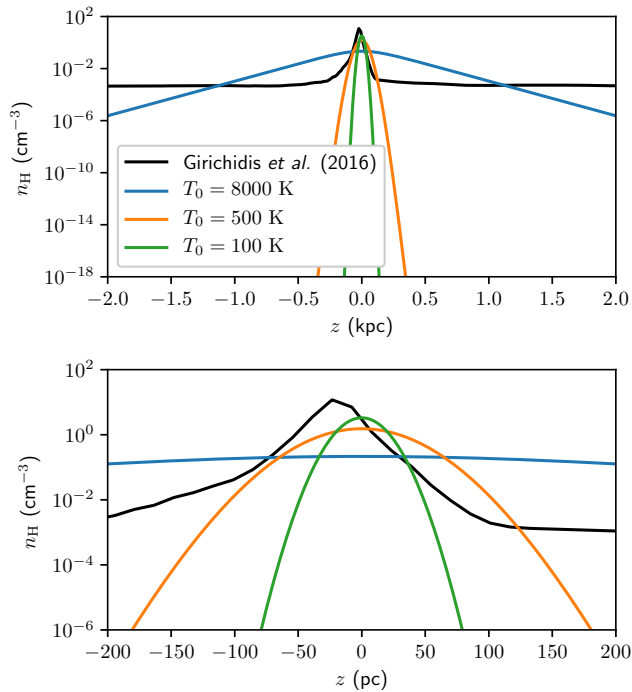}
\caption{Equilibrium hydrostatic density profile for three different values of
the equilibrium temperature $T_0$ (as indicated in the legend) and a fixed gas
surface density $\Sigma{}_g = 2~{\rm{}M}_\odot{}~{\rm{}pc}^{-2}$.
For reference, we
also plot the average density for one of the simulations with only thermal
feedback of \citet{2016Girichidis_CR}. The bottom panel shows a zoom on the
central part of the top panel.
\label{figure:initial_density}}
\end{figure}

The initial density profile is plotted in 
\figureref{figure:initial_density} for different values of the initial 
equilibrium temperature $T_0$. Note that for realistic neutral gas 
temperatures $\sim{}100~{\rm{}K}$ the density profile is relatively 
steep and has a lower scale height than found in models that employ a 
detailed supernova feedback model \citep{2016Girichidis_CR}. This 
indicates that detailed feedback modelling provides additional heating 
that increases the average temperature in the disc. For our assumed 
neutral temperature $T_n=500~{\rm{}K}$, we find a central density 
profile that is more similar to a supernova supported disc. Since we do 
not include any other feedback mechanisms apart from photoionization 
feedback, we will use this temperature as our assumed neutral 
temperature in the two-temperature equation of state to mimick the 
dynamical effect of supernova feedback.

For an equilibrium temperature of $8,000~{\rm{}K}$, the scale height of 
the density profile increases significantly. This already indicates that 
the temperature increase caused by photoionization can indeed lead to a 
more extended, diffuse gas component, provided that enough radiation 
makes it out of the disc to ionize out to sufficiently high altitudes. 

Note that the absence of supernova feedback injection will affect the 
amount of turbulence in our disc, and hence the creation of chimneys 
through which ionizing radiation can escape. These chimneys will need to 
be created by photoionization. Our somewhat higher assumed neutral 
temperature should aid this process. Nonetheless, this means we cannot 
expect to constrain the ionizing source luminosity $Q_{\rm{}H}$ with these 
models, and hence treat it as a parameter.

Finally, our choice to start with a fully ionized gas at $T_i = 
8,000~{\rm{}K}$ could also affect our results. We will test the impact 
of this choice by comparing with a simulation that starts with a fully 
neutral gas at $T_n = 500~{\rm{}K}$.

\subsection{Post-processing}

The post-processing step is very similar to the full model used in V18. 
We assume the same abundances for hydrogen, helium and all coolants, 
i.e. ${\rm{}He/H} = 0.1$, ${\rm{}C/H} = 1.4 \times{} 10^{-4}$, 
${\rm{}N/H} = 6.5 \times{} 10^{-5}$, ${\rm{}O/H} = 4.3 \times{} 
10^{-4}$, ${\rm{}Ne/H} = 1.17 \times{} 10^{-4}$ and ${\rm{}S/H} = 1.4 
\times{} 10^{-5}$. We also use the same ionizing source spectrum 
corresponding to a $40,000~{\rm{}K}$ star with a surface gravity of 
$\log{}(g/({\rm{}m~s}^{-1}))=3.40$. However, we now will use the source 
positions and ionizing luminosity $Q_{\rm{}H}$ that was also used for the RHD 
step, for consistency. The RHD step already uses a regular Cartesian 
grid; we will use the same grid for the post-processing step.

Since the resolution of our simulations is similar to that in V18, we 
will use similar Monte Carlo parameters: $10^8$ photon packets for 20 
iterations.

To test the impact of an additional distribution of hot evolved stars on 
our results, we will run two additional post-processing simulations for 
our reference model with an ionising source luminosity of 
$Q_{\rm{}H}=10^{48}~{\rm{}s}^{-1}$. These simulations include an 
additional smooth distribution of hot stars, with a total ionizing 
luminosity of $10^{48}~{\rm{}s}^{-1}$ ($\sim{}5~\%$ of the total 
ionizing budget in the simulation). We assume a flat ionizing spectrum 
between 1 and 4~Ryd, and futhermore assume a Gaussian distribution. 
Model ES1 assumes a scale height of 200~pc for the additional sources 
(consistent with the assumed gravitational potential), while model ES2 
has a scale height of 400~pc. For these simulations, we double the 
number of photon packets; the extra $10^8$ photon packets are used to 
sample the additional component (with a lower weight per photon 
packet).

\section{Results and discussion}

\subsection{Convergence}

\subsubsection{Resolution}

\begin{figure}
\centering{}
\includegraphics[width=0.48\textwidth]{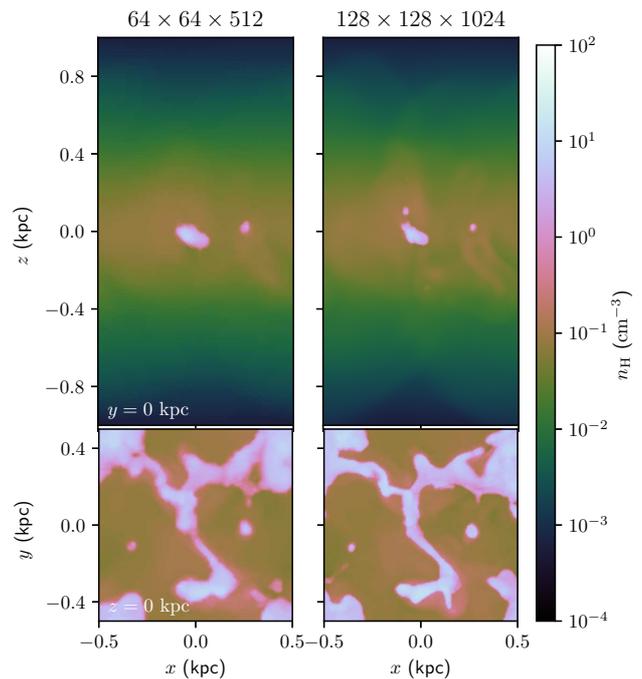}
\caption{Density slices through simulations with two different resolutions, as
indicated in the titles. The top row shows slices along the plane
$y = 0~{\rm{}kpc}$, while the bottom row shows slices along the plane
$z = 0~{\rm{}kpc}$. Both simulations are shown after $200~{\rm{}Myr}$ of
evolution.
\label{figure:convergence_resolution_maps}}
\end{figure}

\begin{figure}
\centering{}
\includegraphics[width=0.48\textwidth]{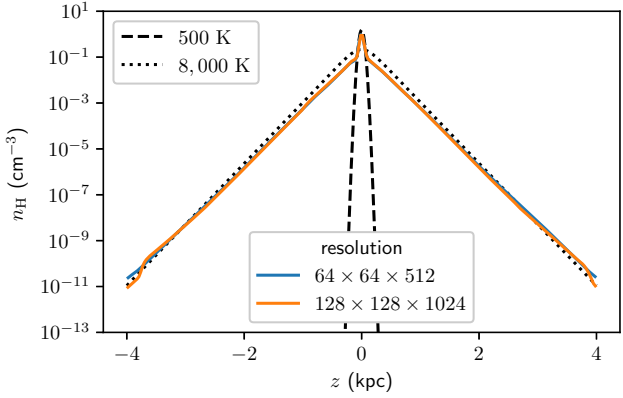}
\caption{Average density as a function of altitude above the disc ($z$) for
simulations with different numerical resolutions, as indicated in the legend.
For reference, we also show the hydrostatic equilibrium curves for the neutral
and ionized temperatures as black curves. Both simulations are shown after
$200~{\rm{}Myr}$ of evolution.
\label{figure:convergence_resolution}}
\end{figure}

Before we can discuss the physical results of our simulations, we have 
to make sure we understand how our adopted numerical resolution affects 
our results. \figureref{figure:convergence_resolution_maps} shows the 
density structure of the disc after $200~{\rm{}Myr}$ of evolution, for 
two simulations that only differ in the numerical resolution that was 
used. The density structures look overall similar, but the higher 
resolution simulation shows more small-scale features that are not resolved 
in the low resolution version.

The average density profiles for both runs are shown in 
\figureref{figure:convergence_resolution}. Overall, the density profiles 
are in good agreement; our low resolution simulations can be considered 
to be converged for our purposes.

\subsubsection{Radiation time step}

\begin{figure}
\centering{}
\includegraphics[width=0.48\textwidth]{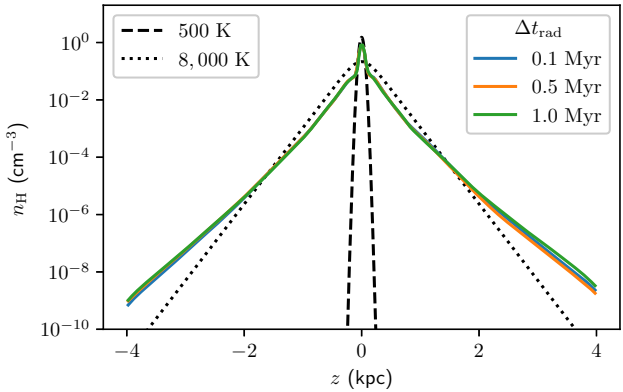}
\caption{Average density as a function of altitude above the disc ($z$) for
simulations with three different values of the radiation field update time
step $\Delta{}t_{\rm{}rad}$, as indicated in the legend. For reference, we
also show the hydrostatic equilibrium curves for the neutral and ionized
temperatures as black curves. All simulations are shown after $800~{\rm{}Myr}$
of evolution.
\label{figure:convergence_radiation}}
\end{figure}

\begin{figure}
\centering{}
\includegraphics[width=0.48\textwidth]{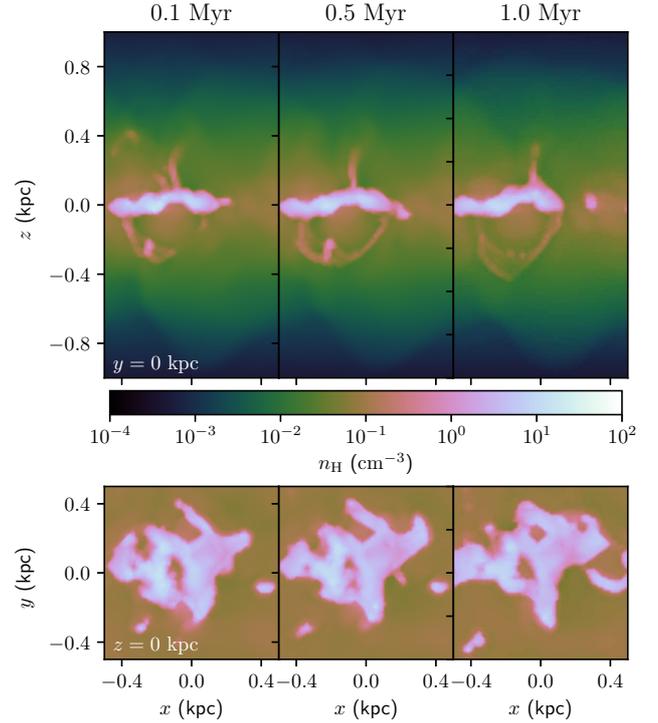}
\caption{Density slices through the simulations with different values of the
radiation time step $\Delta{}t_{\rm{}rad}$, as indicated in the titles. The
top row shows slices along the plane $y=0~{\rm{}kpc}$, while the bottom row
shows slices along the plane $z=0~{\rm{}kpc}$. All simulations are shown after
$800~{\rm{}Myr}$ of evolution.
\label{figure:convergence_radiation_maps}}
\end{figure}

\figureref{figure:number_of_sources} shows the number of discrete 
sources as a function of time for all of our models. The average number 
of sources is $\sim{}23.3$, close to the target number of 24, but there 
is some scatter around this value that happens on short time scales.

This scatter, together with the dynamical effect of the radiation, will 
change the way the radiation interacts with the ISM over time. In 
\ref{subsubsection:radiation} we mentioned that updating the ionization 
state of the gas in the simulation after every hydrodynamical time step 
would be too expensive, and that we only do the radiation step after a 
fixed time interval. Here we check what the effect of this approximation 
is on our final result.

\figureref{figure:convergence_radiation} shows the density profile for 
the three different radiation time steps after $800~{\rm{}Myr}$ of 
evolution. The three profiles are in good agreement over most of the 
range and only diverge in the low density outskirts.

\figureref{figure:convergence_radiation_maps} shows slices through the 
density structure for the same models. These are also in excellent 
agreement; the only noticeable differences are some small scale features 
that are caused by a lack of time resolution for the models with a 
longer radiation time step.

Since we are mainly interested in the large scale structure of the DIG, 
we conclude that using a radiation time step $\Delta{}t_{\rm{}rad} = 
0.5~{\rm{}Myr}$ is sufficient to get representative results.

\subsubsection{Initial gas temperature}

A final model choice that could have a significant impact on our results 
is the initial ionization state and temperature of the gas. As pointed 
out in \sectionref{subsection:initial_conditions}, the neutral 
hydrostatic density profile is very steep, leading to very low values of 
the density at high altitudes above the disc. This leads to numerical 
problems, as the density values reach values very close to the numerical 
precision of a double precision floating point variable. This explains 
our choice for a default ionized initial condition that has a better 
numerical behaviour.

But this choice has profound implications for the dynamics of the 
simulation. For sufficiently low values of the ionizing luminosity per 
source, we expect to obtain a gas with two distinct phases: an ionized 
and a neutral phase. Since the initial condition corresponds to a single 
one of those phases, there will be an initial adaptation phase during 
which the initial gas converts part of its gas into the appropriate 
second phase. This will lead to a large scale relaxation flow in the 
box. If we start with an ionized initial condition, then the relaxation 
flow will be directed inwards: the lack of neutral phase leads to an 
average gas temperature that is lower than the assumed equilibrium and 
cold gas will fall towards the centre of the gravitational potential. 
For a neutral initial condition the reverese happens, and warm gas is 
expelled from the central disc.

\begin{figure}
\centering{}
\includegraphics[width=0.48\textwidth]{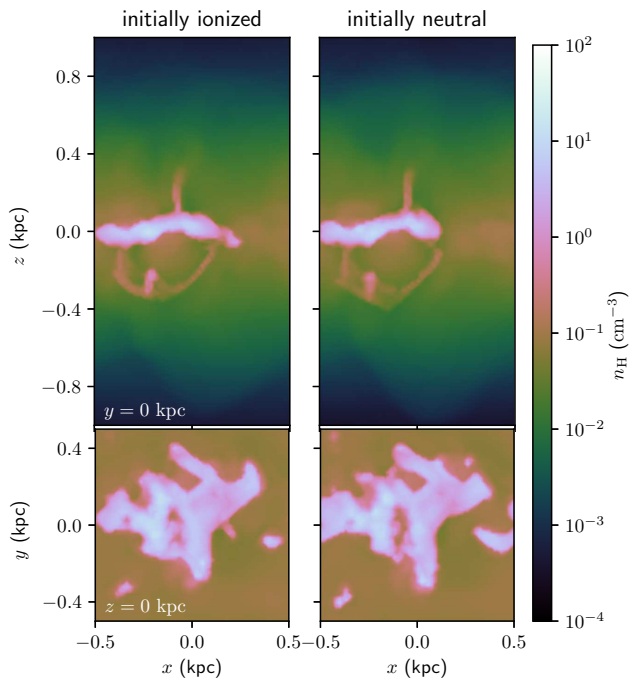}
\caption{Density slices through the simulations with different initial
conditions for the gas, as indicated in the titles. The
top row shows slices along the plane $y=0~{\rm{}kpc}$, while the bottom row
shows slices along the plane $z=0~{\rm{}kpc}$. All simulations are shown after
$800~{\rm{}Myr}$ of evolution.
\label{figure:convergence_initial_condition_maps}}
\end{figure}

\begin{figure}
\centering{}
\includegraphics[width=0.48\textwidth]{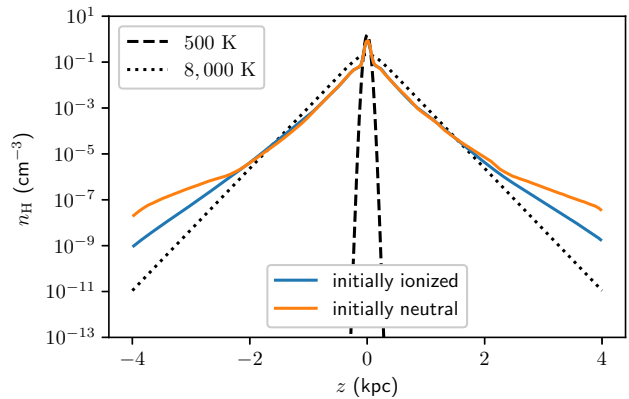}
\caption{Average density as a function of altitude above the disc ($z$) for
simulations with different initial conditions for the gas,
as indicated in the legend. For reference, we also show the hydrostatic
equilibrium curves for the neutral and ionized temperatures as black curves.
All simulations are shown after $800~{\rm{}Myr}$ of evolution.
\label{figure:convergence_initial_condition}}
\end{figure}

\figureref{figure:convergence_initial_condition_maps} shows the density 
structure for the two different initial condition scenarios after 
$800~{\rm{}Myr}$ of evolution. The two simulations are in good 
agreement, illustrating that by this time, a steady-state equilibrium is 
reached for the central part of the disc that is independent of the 
initial condition. This can also be seen in the average density profiles 
in \figureref{figure:convergence_initial_condition}~: the central 
density profiles agree very well. The difference in large scale 
relaxation flow is however still apparent from the average density at 
higher altitudes, which is significantly higher for the neutral initial 
condition, consistent with a relaxation flow that is directed outwards.

\subsection{Ionizing luminosity}
\label{subsection:ionizing_luminosity}

\begin{figure}
\centering{}
\includegraphics[width=0.48\textwidth]{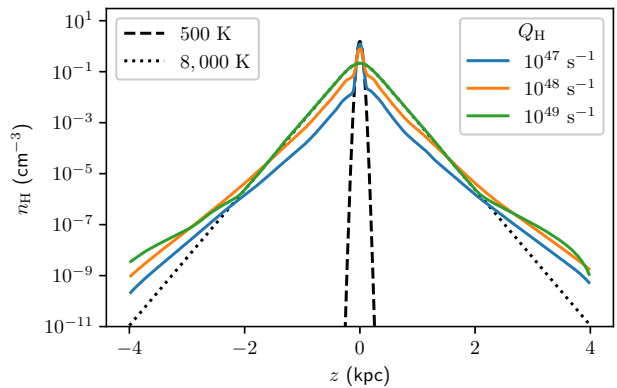}
\caption{Average density as a function of altitude above the disc ($z$) for
simulations with three different values of the ionizing luminosity $Q_{\rm{}H}$,
as indicated in the legend. For reference, we also show the hydrostatic
equilibrium curves for the neutral and ionized temperatures as black curves.
All simulations are shown after $800~{\rm{}Myr}$ of evolution.
\label{figure:convergence_luminosity}}
\end{figure}

For this work, we will assume that our external potential and equation 
of state are fixed, so that our model is entirely determined by the 
parameters of the ionizing source model. Since the number of ionizing 
sources and their scale height has been chosen based on observations, 
there is only one free parameter in our model: the ionizing luminosity 
for an individual source, $Q_{\rm{}H}$.

\begin{figure}
\centering{}
\includegraphics[width=0.48\textwidth]{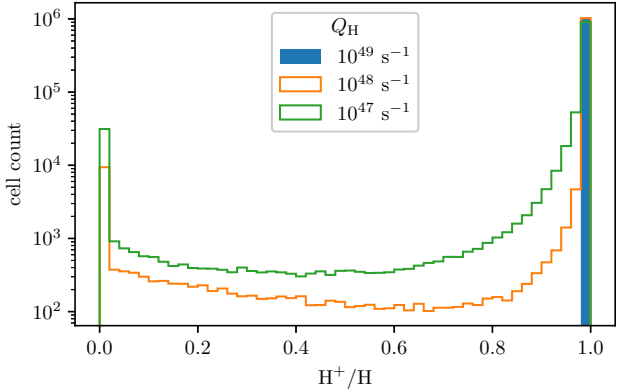}
\caption{Histograms of the ionized gas fraction for
simulations with three different values of the ionizing luminosity $Q_{\rm{}H}$,
as indicated in the legend. Only the cells with $|z| < 2~{\rm{}kpc}$ are
shown, since cells at higher altitudes only contribute to the lowest bin.
All simulations are shown after $800~{\rm{}Myr}$ of evolution.
\label{figure:luminosity_histogram}}
\end{figure}

\figureref{figure:convergence_luminosity} shows the average density as a 
function of altitude above the disc for three different values of the 
ionizing luminosity $Q_{\rm{}H}$. It is clear that higher values of the 
ionizing luminosity lead to overall higher average densities for the 
diffuse gas, and lower densities for the neutral disc, as gas is 
converted from being in a quasi-hydrostatic neutral state into a 
quasi-hydrostatic ionized state. This is even more evident from the 
histograms shown in \figureref{figure:luminosity_histogram} that show 
the number of cells with a given neutral fraction for hydrogen, 
$x_{\rm{}H}$: as the ionizing luminosity increases, the number of cells 
within the (partially) neutral bins decreases, and more cells enter the 
highly ionized bin. Note also the clear separation between the ionized 
and the neutral state in these histograms.

That the gas is not actually in strict hydrostatic equilibrium is clear 
from the tails of the average density profiles: all of the models have 
an extended tail that lies above the expected hydrostatic equilibrium 
curve. This corresponds to gas that is either outflowing or is falling 
back to the centre of the gravitational potential after being previously 
expelled. This behaviour is to be expected, as the source model itself 
is constantly evolving and causing small-scale movements of the gas.

\begin{figure}
\centering{}
\includegraphics[width=0.48\textwidth]{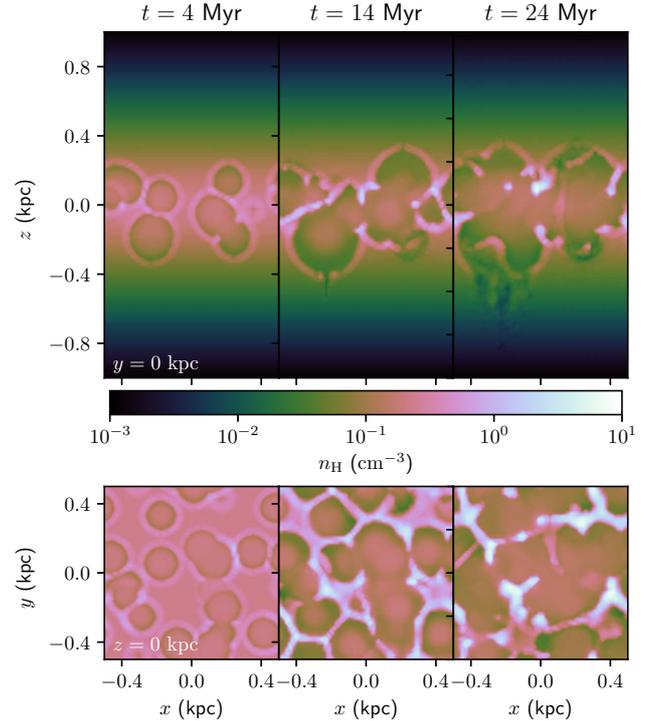}
\caption{Three snapshots of the time evolution of the ionized bubbles in the
early phases of the same simulation, showing the compression of the gas in
regions where multiple bubbles collide. The simulation shown is the low
resolution reference model with an ionizing luminosity
$Q_{\rm{}H} = 10^{48}~{\rm{}s}^{-1}$ and radiation time step
$\Delta{}t_{\rm{}rad} = 0.5~{\rm{}Myr}$.
\label{figure:bubble_expansion}}
\end{figure}

There are a number of explanations for the increasing conversion from 
neutral gas in the central disc to ionized gas in the diffuse disc with 
increasing ionizing luminosity. As can be seen from the early evolution 
of one of our models in \figureref{figure:bubble_expansion}, the central 
density structure is predominantly governed by the interaction of 
expanding ionization bubbles surrounding the individual ionizing 
sources. These bubbles collide as they expand and gas in the resulting 
shock fronts is compressed into dense structures that are hard to 
ionize. At the same time, the part of the ionizing shell that expands in 
the vertical direction carries dense gas with it and expels it into the 
diffuse ionized layer. Higher ionizing luminosities lead to ionization 
bubbles that expand faster and hence more expelled material and denser 
structures in the disc plane. In the extreme case of a very high 
ionizing luminosity, the dense structures in the disc plane can still be 
ionized effectively and no neutral gas is left.

\subsection{Line emission}

Thus far, we have only discussed the density structures that result from 
our RHD simulations. We will now post-process the models with different 
values for the ionizing luminosity using the same model as was used for 
V18, to compute a more detailed temperature structure and line emission 
strengths. Note that we use the same source positions and ionizing 
luminosities that were used during the RHD simulation, so that our model 
is self-consistent. This is a significant improvement over V18, where 
the source positions were chosen randomly and had no link to the spatial 
locations of feedback events in the underlying SILCC model. Also recall 
that these models are not intended to provide a full spectral model for 
the DIG; we are only quantifying the impact of ionizing radiation from 
O/B stars in the disc on the spectral properties of the DIG.

\begin{figure}
\centering{}
\includegraphics[width=0.48\textwidth]{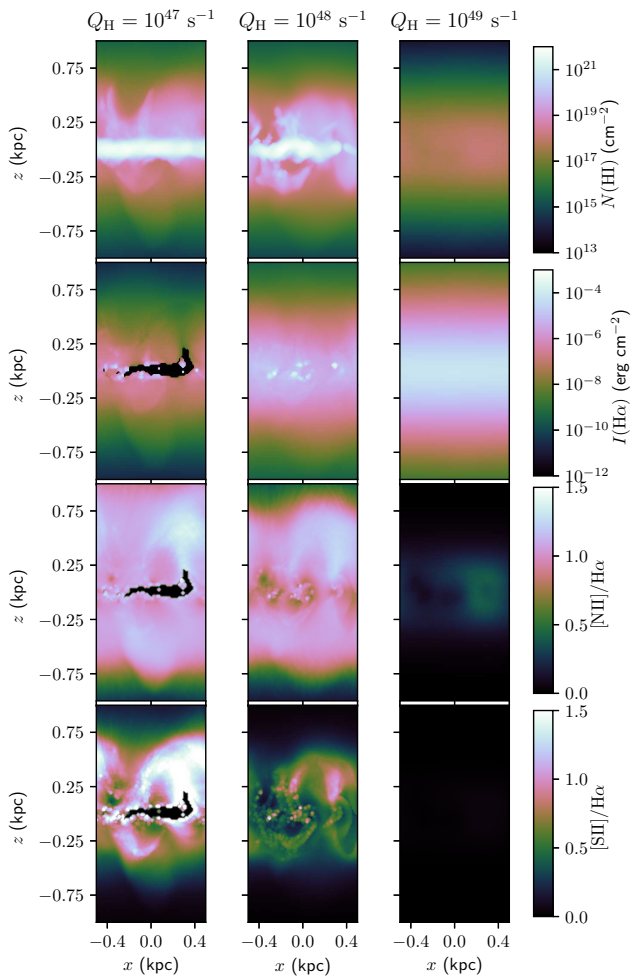}
\caption{Line emission maps for (from top to bottom): neutral hydrogen,
${\rm{}H}\alpha{}$,  $[{\rm{}NII}]/{\rm{}H}\alpha{}$ and 
$[{\rm{}SII}]/{\rm{}H}\alpha{}$ for the low resolution simulations with
different values of the ionizing luminosity.
All simulation results were obtained by
post-processing the output of the different simulations at $t=800~{\rm{}Myr}$,
as shown in \figureref{figure:convergence_luminosity}.
\label{figure:all_maps}}
\end{figure}

\figureref{figure:all_maps} shows line emission maps for the main 
tracers of the neutral and ionised gas in our discs for different 
ionizing source luminosities. All of these were created by integrating 
the contributions of individual cells along the $y$-direction. It is 
immediately clear that the highest luminosity result lacks any clear 
neutral disc. The neutral disc in the other two simulations shows clear 
structures of outflows driven by expanding and contracting 
photoionization bubbles which look similar to observed Milky Way 
features \citep{1984Heiles, 1992Koo, 2012Hartmann}, as well as clear 
signatures of active \textsc{Hii} regions in the disc (these show up as 
bright spots in ${\rm{}H}\alpha{}$), similar to e.g. the Galactic maps 
from the WHAM survey \citep{2003Haffner}. For the simulations with a 
neutral disc, we see a clear trend in the 
$[{\rm{}NII}]/{\rm{}H}\alpha{}$ and $[{\rm{}SII}]/{\rm{}H}\alpha{}$ line 
ratios: these tend to increase towards higher altitudes above the disc.

\begin{figure}
\centering{}
\includegraphics[width=0.48\textwidth]{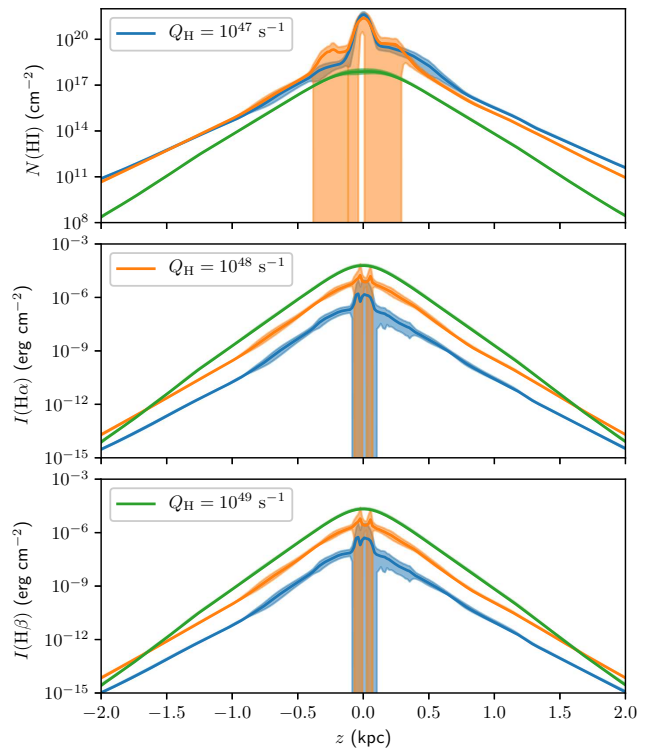}
\caption{Profiles of the neutral hydrogen, H$\alpha{}$ intensity, and H$\beta{}$
intensity as a function
of height above the disc
for the low resolution simulations with different values of the ionizing
luminosity, as indicated in the legend. The full lines correspond to the average
values for lines of equal $z$ in \figureref{figure:all_maps},
while the shaded regions represent the standard
deviation over the same lines (these are asymmetric on the panels with a
logarithmic $y$-axis).
All simulation
results were obtained by post-processing the output of the different simulations
at $t=800~{\rm{}Myr}$, as shown in \figureref{figure:convergence_luminosity}.
\label{figure:compare_Halpha}}
\end{figure}

\begin{figure}
\centering{}
\includegraphics[width=0.48\textwidth]{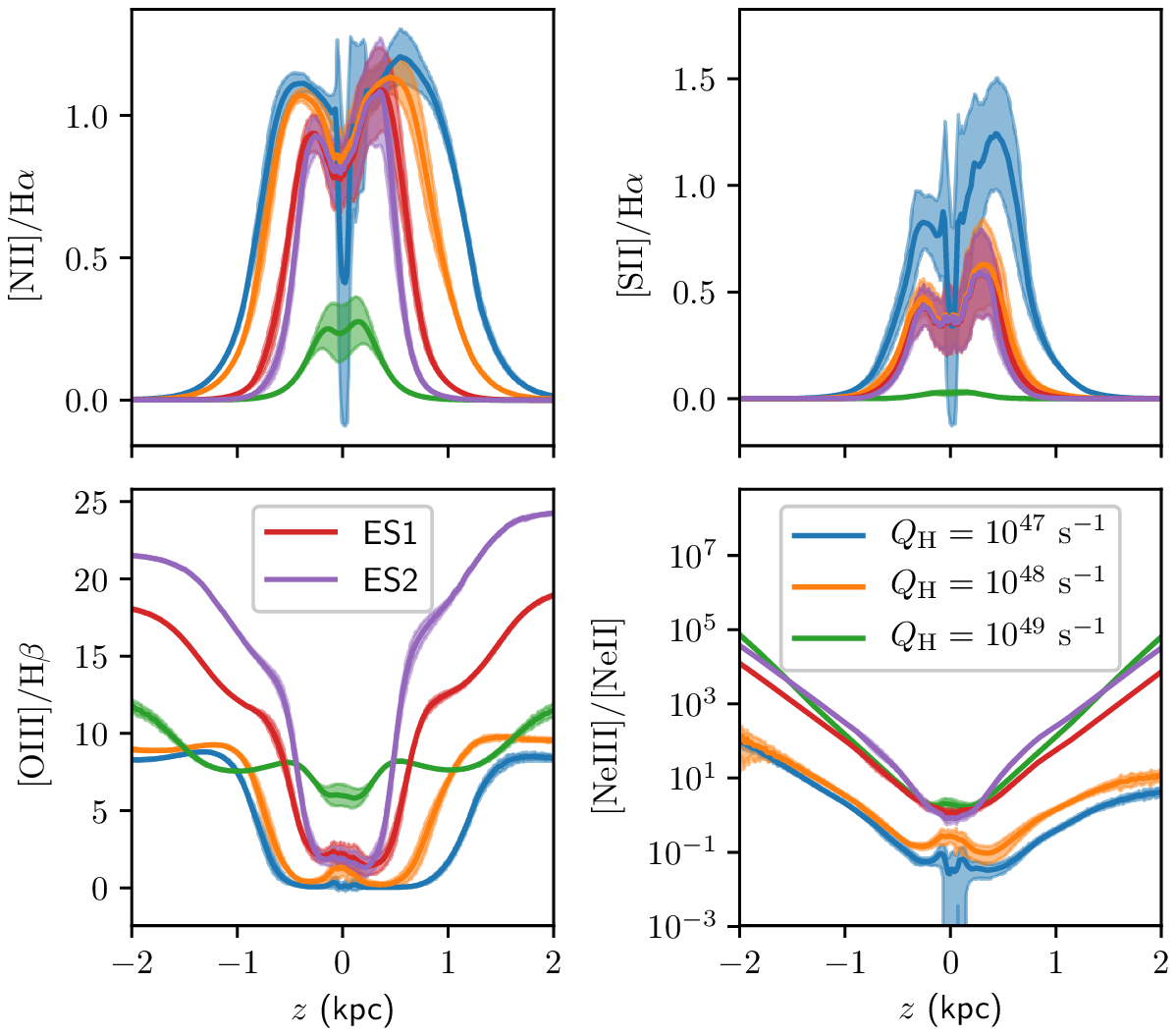}
\caption{Line ratios for $[{\rm{}NII}]/{\rm{}H}\alpha{}$,
$[{\rm{}SII}]/{\rm{}H}\alpha{}$, $[{\rm{}OIII}]/{\rm{}H}\beta{}$ and
$[{\rm{}NeIII}]/[{\rm{}NeII}]$ as a function of height above the plane of the
disc, for the same simulations shown in \figureref{figure:compare_Halpha}.
We also show the results for two additional models (ES1 and ES2) that use
a source luminosity $Q_{\rm{}H}=10^{48}~{\rm{}s}^{-1}$ and assume an additional
hot evolved stellar component with an ionising luminosity of
$10^{48}~{\rm{}s}^{-1}$ with a scale height of 200~pc and 400~pc.
\label{figure:line_ratios_all}}
\end{figure}

The corresponding emission profiles for all simulations are shown in 
\multifigureref{figure:compare_Halpha}{figure:line_ratios_all}. These 
profiles were created by averaging the emission maps in 
\figureref{figure:all_maps} over all $x$-directions. The H$\alpha{}$ 
emission increases with increasing ionizing luminosity, while the 
neutral hydrogen has the opposite trend. Note that the H$\alpha{}$ 
emission is relatively low compared to that in the observed DIG, 
indicating that our diffuse gas densities are too low. This might be 
explained by the absence of strong supernova feedback in our 
simulations, or by the fact that we do not take into account the 
fluorescence of the Lyman lines \citep{2011FloresFajardo}. Also note 
that the high luminosity curve has almost no scatter, indicative of the 
fact that all the gas in this simulation is highly ionized.

\figureref{figure:line_ratios_all} also shows the impact of an 
additional hot evolved star component in our 
$Q_{\rm{}H}=10^{48}~{\rm{}s}^{-1}$ reference model. This additional 
component has a negligible impact on the neutral gas fractions and 
${\rm{}H}\alpha{}$ and ${\rm{}H}\beta{}$ emission, and very little 
impact on the $[{\rm{}SII}]/{\rm{}H}\alpha{}$ ratio. It does lead to a 
noticeable shift in the $[{\rm{}NII}]/{\rm{}H}\alpha{}$ ratio at higher 
altitudes, indicative of a shift from ${\rm{}N}^{+}$ to ${\rm{}N}^{++}$ 
as dominant nitrogen ion. This is mirrored in the ionization structure 
of oxygen, with a clear increase of the 
$[{\rm{}OIII}~4959+5007~$\AA{}$]/{\rm{}H}\beta{}$ line ratio. The line 
ratio of 
$[{\rm{}NeIII}~15~\mu{}{\rm{}m}]/[{\rm{}NeII}~12~\mu{}{\rm{}m}]$ is 
affected most by the additional evolved star component: while the 
original models have a slowly increasing ratio as a function of height, 
consistent with the \citet{2011Rand} observations, the additional 
component boosts this ratio by two orders of magnitude. This is 
consistent with the results for neon found in \citet{2011FloresFajardo}.

\begin{figure}
\centering{}
\includegraphics[width=0.48\textwidth]{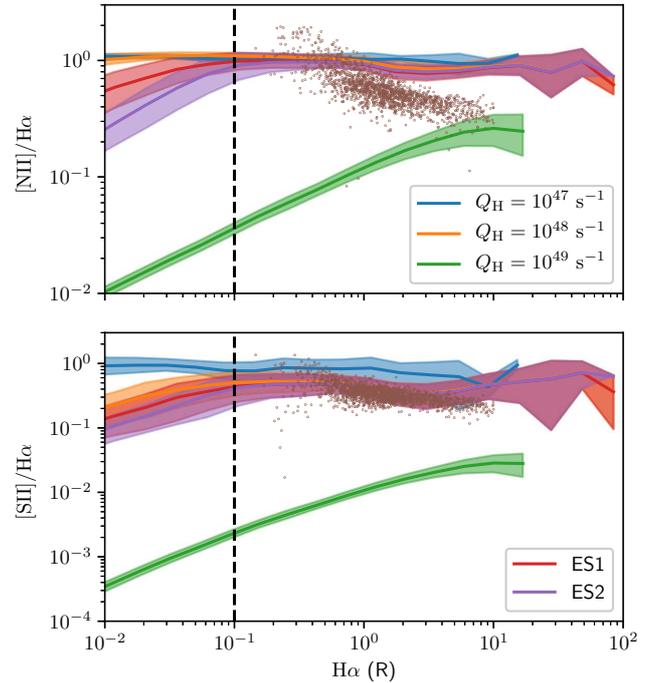}
\caption{Line ratios of the forbidden nitrogen and sulphur lines
as a function of the H$\alpha{}$ emission for our low resolution models
with different values for the ionizing luminosity and the two additional models
with a hot evolved star component, as indicated in the
legend. The full lines denote the average line ratio within 50 logarithmic
bins, while the shaded regions represent the standard deviation within these
bins. The dashed line is the sensitivity limit of the Wisconsin H$\alpha{}$
mapper and is indicative for the range that is accessible in observations.
The dots represent reference observational data from the WHAM survey
\citep{1999Haffner}. ${\rm{}H}\alpha{}$ intensities are quoted in Rayleigh
to facilitate comparison with the WHAM observations.
All simulation results were obtained by post-processing
the output of the different simulations at $t=800~{\rm{}Myr}$, as shown in
\figureref{figure:convergence_luminosity}.
\label{figure:compare_lines}}
\end{figure}

\figureref{figure:compare_lines} shows the relative strength of the 
forbidden line emission from [N\textsc{ii}]~6584~\AA{} and 
[S\textsc{ii}]~6725~\AA{} as a function of the H$\alpha{}$ emission 
strength. The models with low luminosities reproduce the observed 
sulphur line ratio trend; the intermediate luminosity even matches the 
observed values. The high luminosity result has much lower line ratios 
than observed, again indicative of its unrealistic ionization structure. 
For the nitrogen line ratios the correspondence is less, but we can 
still see a similar trend in the low luminosity results. It is important 
to note that our models ignore Ly$\alpha{}$ fluorescence and therefore 
underestimate the total ${\rm{}H}\alpha{}$ luminosity.

The line ratios still depend on the ionizing luminosity, but unlike in 
V18, this now can be linked to the dynamic effect of photoionization: 
luminosities that are too high ionize too much gas. This does not only 
lead to line ratios that are lower than observed, but also to a DIG 
structure that is denser than observed, and the complete absence of a 
neutral disc. As in V18, luminosities that are too low lead to realistic 
line ratios, but fail to support any significant DIG. We conclude that 
the observed line ratios are indicative of effective spectral hardening 
throughout the DIG (as can be seen in the temperature profiles in V18), 
and that such a signature is a natural consequence of the DIG being 
supported by the increase in temperature caused by photoionization. 
Furthermore, our O/B star only model agrees well with observed line 
ratio trends; an additional hot evolved star component with a much 
harder spectrum has little impact, and fails to reproduce the observed 
neon line ratio.

\section{Conclusion}

In this work, we have presented the first self-consistent RHD 
simulations of the DIG in disc galaxies. We show that a fully 
self-consistent treatment of photoionization as part of the dynamical 
modelling of a disc galaxy slab naturally leads to a diffuse ionized 
disc component that is supported by photoionization heating. This layer 
has a temperature structure that is in line with observed DIG 
properties, provided that the ionizing luminosity is sufficiently strong 
to support a DIG, and weak enough to allow for the existence of a 
neutral disc. The fine-tuning found in previous work (V18) is hence 
naturally explained by the dynamic impact of photoionizing radiation.

Despite being very basic, our models are relatively expensive, forcing 
us to make strong assumptions about the coupling between radiation and 
hydrodynamics. This means that we cannot currently constrain the 
properties of the ionizing radiation field, nor make strong predictions 
about the actual temperature structure of the DIG that would result from 
an observed distribution of sources, nor provide a full spectral model 
of the DIG that explains all observed line ratios. Nevertheless, our 
models provide a valuable insight into the close link between the 
dynamical effect of photoionizing radiation and the observed properties 
of the DIG.

The diffuse gas in our models is generally more diffuse than that in the 
observed DIG. This indicates that (a) the DIG is not in hydrostatic 
equilibrium, as our densities are close to the expected hydrostatic 
equilibrium values, but is instead either outflowing or being accreted, 
(b) we are missing additional feedback energy that provides extra 
support or drives strong outflows that boost the density in the diffuse 
component. Supernova feedback is one clear missing feedback mechanism 
\citep{2015Walch}; additional mechanisms like cosmic ray feedback 
\citep{2016Girichidis_CR} or magnetic fields \citep{2012Hill} can be 
included as well. A full quantitative model for the DIG should also 
contain more realistic stellar luminosities, lifetimes and spectra, and 
a more physically motivated source distribution model that places UV 
sources in dense neutral clumps and that includes photoionizing 
radiation from hot evolved sources. All of these can be addressed in 
future work to provide a more realistic model that will enable more 
detailed comparisons with observations.

\section*{Acknowledgements}

We thank Alex S. Hill for comments that improved the quality of this 
work. BV and KW acknowledge support from STFC grant ST/M001296/1. Part 
of this work was performed using the DiRAC Data Intensive service at 
Leicester, operated by the University of Leicester IT Services, which 
forms part of the STFC DiRAC HPC Facility (www.dirac.ac.uk). The 
equipment was funded by BEIS capital funding via STFC capital grants 
ST/K000373/1 and ST/R002363/1 and STFC DiRAC Operations grant 
ST/R001014/1. DiRAC is part of the National e-Infrastructure.

\bibliographystyle{mnras}
\bibliography{main}

\bsp    
\label{lastpage}

\end{document}